\documentclass[12pt]{article}
\usepackage{graphicx}
\usepackage{amsmath}
\usepackage{amssymb}
\usepackage{caption2}
\begin{document}
\bibliographystyle{prsty}
\begin{center}
{\large {\bf \sc{   Axial  form-factor and induced pseudoscalar form-factor of the nucleons }}} \\[2mm]
Zhi-Gang Wang$^{1}$ \footnote{Corresponding author; E-mail,wangzgyiti@yahoo.com.cn.  }, Shao-Long Wan$^{2}$ and Wei-Min Yang$^{2} $    \\
$^{1}$ Department of Physics, North China Electric Power University, Baoding 071003, P. R. China \\
$^{2}$ Department of Modern Physics, University of Science and Technology of China, Hefei 230026, P. R. China \\
\end{center}

\begin{abstract}
In this article, we  calculate the
  axial and the induced pseudoscalar form-factors $G_A(t=-Q^2)$ and $G_P(t=-Q^2)$ of the nucleons in the
  framework of the light-cone QCD sum-rules approach up
 to twist-6 three valence quark light-cone distribution amplitudes,
  and observe that the  form-factors $G_A(t=-Q^2)$ and $G_P(t=-Q^2)$ at intermediate and
large momentum transfers with $Q^2> 2 GeV^2$ have significant
contributions from the end-point (soft) terms.  The numerical values
for the axial form-factor $G_A(t=-Q^2)$ are
 compatible with the experimental data and  theoretical calculations, for
 example, the chiral quark models and lattice QCD.
 The numerical values for the induced pseudoscalar form-factor
 $G_P(t=-Q^2)$ are compatible with the calculation from the
 Bethe-Salpeter equation.
 \end{abstract}

PACS : 12.38.Aw, 12.38.-t, 14.20.Dh

{\bf{Key Words:}}  Axial-vector current, Light-cone QCD sum rules,
Form-factor
\section{Introduction}
The  axial and induced pseudoscalar form-factors of the nucleons are
of fundamental importance in studying the weak interactions and the
pion-nucleon scattering. They provide  an important test for
theories which attempt to describe the under-structures  of the
nucleons and the underlying dynamics \cite{AdlerBook,Review2002}.
Using Lorentz covariance and chiral symmetry, the matrix element of
the axial-vector current between the initial and final  nucleon
states excluding the second class current \cite{Weinberg58} can be
parameterized as,
\begin{eqnarray}
\langle N(P')|A_\mu^a(0)|N(P)\rangle=N(P')\left\{\gamma_\mu
G_A(t)+\frac{(P'-P)_\mu}{2M}G_P(t) \right\}\gamma_5\tau^a N(P),
\end{eqnarray}
here the $\tau^a$ is the Pauli matrix, the $M$ is the average  mass
for the proton and neutron,  $t=(P'-P)^2$, the $G_A(t)$ and $G_P(t)$
are the axial and induced pseudoscalar form-factor respectively.
 The Goldberger-Treiman relation \cite{GT58} relates the
 form factors  $G_A(t)$ and $G_P(t)$, and
  the pion decay constant $f_{\pi}$,
\begin{eqnarray}
g_{\pi NN}={g_A M\over f_\pi}, \ \ \ g_{\pi NN}=G_P(t=m_\pi^2), \ \
\ g_A=G_A(t=0) \, .
\end{eqnarray}

In this article, we calculate the axial form-factor $G_A(t)$ and
induced pseudoscalar form-factor $G_P(t)$ of   the nucleons in the
framework of the light-cone sum rules (LCSR) approach \cite{LCSR,
LCSRreview} which combine the standard techniques  of the QCD sum
rules with the conventional parton distribution amplitudes
describing  the hard exclusive processes\cite{SVZ79}. In the LCSR
approach, the short-distance operator product expansion with   the
vacuum condensates of increasing dimensions is replaced by the
light-cone expansion with the distribution amplitudes (which
correspond to the sum of an infinite series of operators with the
same twist) of increasing twists  to parameterize the
non-perturbative  QCD vacuum, while the contributions from  the hard
re-scattering can be correctly incorporated  as the $O(\alpha_s)$
corrections \cite{BKM00}. In recent years, there have been a lot of
applications of the LCSR to the mesons, for example, the
form-factors, strong coupling constants and hadronic matrix elements
\cite{LCSRreview}, the applications to the baryons are cumbersome
 and only the  electromagnetic form factors
 \cite{BaryonBraun}, the scalar form-factor \cite{WangWY} and the weak decay $\Lambda_b\to
p\ell\nu_\ell$  \cite{Huang04} are studied,  the higher twists
distribution amplitudes   of the baryons were not available until
recently \cite{BFMS}.

The article is arranged as follows:  we derive the light-cone sum
rules for the axial and induced pseudoscalar form-factors $G_A(t)$
and $G_P(t)$ of  the nucleons in section II; in section III,
numerical results and discussion; section VI is reserved for
conclusion.

\section{Light-cone sum rules for the  form-factors $G_A(t)$ and $G_P(t)$}
In the following, we write down  the two-point correlation function
$\Pi_\mu(P,q)$ in the framework of the LCSR approach,
\begin{eqnarray}
 \Pi_\mu(P,q) &=& i \int d^4 x \, e^{i q \cdot x}
\langle 0| T\left\{\eta(0) J_\mu (x)\right\} |P\rangle ,
\end{eqnarray}
 with the axial-vector current
\begin{eqnarray}
 J_\mu(x)&=&  \bar{d}(x)\gamma_\mu \gamma_5  u(x)  ,
\end{eqnarray}
 and the neutron  current
\cite{Che84}
\begin{eqnarray}
 \eta(0) &=& \epsilon^{ijk}
\left[d^i(0) C \!\not\!{z} d^j(0)\right] \, \gamma_5 \!\not\!{z}
u^k(0) \,,
\nonumber\\
 \langle0| \eta(0)  |P\rangle & = & f_{\rm N}\,
(P \cdot z) \!\not\!{z} N(P)\, ,
\end{eqnarray}
 here the $z$ is a light-cone vector, $z^2 = 0$, and the  $f_N$ is the
 coupling constant of
  the leading twist light-cone distribution amplitude
\cite{earlybaryon}.  At the large Euclidean momenta ${P'}^2=(P-q)^2$
and $q^2 = - Q^2$, the correlation function $\Pi_\mu(P,q) $ can be
calculated in perturbation theory. In calculation, we need the
following light-cone expanded quark propagator \cite{BB89},
\begin{eqnarray}
S(x) &=& \frac{i\Gamma(d/2)\not\!x}{2\pi^2(-x^2)^{d/2}} \nonumber
\\
&+& \frac{i\Gamma(d/2-1)}{16\pi^2(-x^2)^{d/2-1}}\int\limits_0^1dv
\Big\{(1-v)\not\!x \sigma_{\mu \nu} G^{\mu\nu}(vx) + v\sigma_{\mu
\nu} G^{\mu\nu}(vx)\!\not\! x \Big\} + ...,
 \end{eqnarray} where $
G_{\mu\nu}= g_sG_{\mu\nu}^a(\lambda^a/2)$ is the gluon field
strength tensor and $d$ is the space-time dimension.  The
contributions proportional to the $G_{\mu\nu}$ can give rise to
four-particle (and five-particle) nucleon distribution amplitudes
with a gluon (or quark-antiquark pair) in addition to the three
valence quarks, their corrections are usually not expected to play
any significant roles \cite{DFJK} and neglected here
\cite{BaryonBraun,Huang04}. In the parton model, at large momentum
transfers, the electromagnetic and weak currents interact with the
almost free partons in the nucleons. Employ the "free" light-cone
quark propagator in the correlation function $\Pi_\mu(P,q)$, we
obtain
\begin{eqnarray}
 z^\mu \Pi_\mu(P,q) &=& -2 \int d^4 x \, \frac{z\cdot x e^{i q \cdot x}}{ \pi^2 x^4}
(\gamma_5 \!\not\!{z})^{\lambda \alpha}
(C\!\not\!{z}\gamma_5)^{\beta\gamma} \epsilon^{ijk}\langle 0|
T\left\{ u^i_\alpha(0) u^j_\beta(x)d^k_\gamma(0)\right\} |P\rangle.
\nonumber
\\
\end{eqnarray}

In the light-cone limit $x^2\to 0$, the remaining three-quark
operator sandwiched between the proton state and the vacuum can be
written in terms of the  nucleon distribution amplitudes
\cite{Che84,BFMS,earlybaryon}.  The three valence quark components
of the nucleon distribution amplitudes are defined by the matrix
element,
\begin{eqnarray}
&&4\langle0|\epsilon^{ijk}u_\alpha^i(a_1 x)u_\beta^j(a_2
x)u_\gamma^k(a_3
x)|P\rangle=\mathcal{S}_1MC_{\alpha\beta}(\gamma_5N)_{\gamma}+
\mathcal{S}_2M^2C_{\alpha\beta}(\rlap/x\gamma_5N)_{\gamma}\nonumber\\
&&{}+ \mathcal{P}_1M(\gamma_5C)_{\alpha\beta}N_{\gamma}+
\mathcal{P}_2M^2(\gamma_5C)_{\alpha\beta}(\rlap/xN)_{\gamma}+
(\mathcal{V}_1+\frac{x^2M^2}{4}\mathcal{V}_1^M)(\rlap/PC)_{\alpha\beta}(\gamma_5N)_{\gamma}
\nonumber\\&&{}+
\mathcal{V}_2M(\rlap/PC)_{\alpha\beta}(\rlap/x\gamma_5N)_{\gamma}+
\mathcal{V}_3M(\gamma_\mu
C)_{\alpha\beta}(\gamma^\mu\gamma_5N)_{\gamma}+
\mathcal{V}_4M^2(\rlap/xC)_{\alpha\beta}(\gamma_5N)_{\gamma}\nonumber\\&&{}+
\mathcal{V}_5M^2(\gamma_\mu
C)_{\alpha\beta}(i\sigma^{\mu\nu}x_\nu\gamma_5N)_{\gamma} +
\mathcal{V}_6M^3(\rlap/xC)_{\alpha\beta}(\rlap/x\gamma_5N)_{\gamma}\nonumber\\&&{}
+(\mathcal{A}_1+\frac{x^2M^2}{4}\mathcal{A}_1^M)(\rlap/P\gamma_5
C)_{\alpha\beta}N_{\gamma}+
\mathcal{A}_2M(\rlap/P\gamma_5C)_{\alpha\beta}(\rlap/xN)_{\gamma}+
\mathcal{A}_3M(\gamma_\mu\gamma_5 C)_{\alpha\beta}(\gamma^\mu
N)_{\gamma}\nonumber\\&&{}+
\mathcal{A}_4M^2(\rlap/x\gamma_5C)_{\alpha\beta}N_{\gamma}+
\mathcal{A}_5M^2(\gamma_\mu\gamma_5
C)_{\alpha\beta}(i\sigma^{\mu\nu}x_\nu N)_{\gamma}+
\mathcal{A}_6M^3(\rlap/x\gamma_5C)_{\alpha\beta}(\rlap/x
N)_{\gamma}\nonumber\\&&{}+(\mathcal{T}_1+\frac{x^2M^2}{4}\mathcal{T}_1^M)(P^\nu
i\sigma_{\mu\nu}C)_{\alpha\beta}(\gamma^\mu\gamma_5
N)_{\gamma}+\mathcal{T}_2M(x^\mu P^\nu
i\sigma_{\mu\nu}C)_{\alpha\beta}(\gamma_5
N)_{\gamma}\nonumber\\&&{}+
\mathcal{T}_3M(\sigma_{\mu\nu}C)_{\alpha\beta}(\sigma^{\mu\nu}\gamma_5
N)_{\gamma}+
\mathcal{T}_4M(P^\nu\sigma_{\mu\nu}C)_{\alpha\beta}(\sigma^{\mu\rho}x_\rho\gamma_5
N)_{\gamma}\nonumber\\&&{}+ \mathcal{T}_5M^2(x^\nu
i\sigma_{\mu\nu}C)_{\alpha\beta}(\gamma^\mu\gamma_5 N)_{\gamma}+
\mathcal{T}_6M^2(x^\mu P^\nu
i\sigma_{\mu\nu}C)_{\alpha\beta}(\rlap/x\gamma_5
N)_{\gamma}\nonumber\\&&{}+
\mathcal{T}_7M^2(\sigma_{\mu\nu}C)_{\alpha\beta}(\sigma^{\mu\nu}\rlap/x\gamma_5
N)_{\gamma}+
\mathcal{T}_8M^3(x^\nu\sigma_{\mu\nu}C)_{\alpha\beta}(\sigma^{\mu\rho}x_\rho\gamma_5
N)_{\gamma} \, \, .
\end{eqnarray}
The calligraphic distribution amplitudes do not have definite twist
and can be related to the ones with definite twist as
\begin{eqnarray}
&&\mathcal{S}_1=S_1, \hspace{0.8cm}2P\cdot
x\mathcal{S}_2=S_1-S_2,\nonumber\\&& \mathcal{P}_1=P_1,
\hspace{0.8cm}2P\cdot x\mathcal{P}_2=P_1-P_2 \nonumber
\end{eqnarray}
for the scalar and pseudoscalar distribution amplitudes,
\begin{eqnarray}
&&\mathcal{V}_1=V_1, \hspace{2.4cm}2P\cdot
x\mathcal{V}_2=V_1-V_2-V_3, \nonumber\\&& 2\mathcal{V}_3=V_3,
\hspace{2.2cm} 4P\cdot
x\mathcal{V}_4=-2V_1+V_3+V_4+2V_5,\nonumber\\&& 4P\cdot
x\mathcal{V}_5=V_4-V_3,\hspace{0.5cm} (2P\cdot
x)^2\mathcal{V}_6=-V_1+V_2+V_3+V_4+V_5-V_6 \nonumber
\end{eqnarray}
for the vector distribution amplitudes,
\begin{eqnarray}
&&\mathcal{A}_1=A_1, \hspace{2.4cm}2P\cdot
x\mathcal{A}_2=-A_1+A_2-A_3, \nonumber\\&& 2\mathcal{A}_3=A_3,
\hspace{2.2cm}4P\cdot x\mathcal{A}_4=-2A_1-A_3-A_4+2A_5,
\nonumber\\&& 4P\cdot x\mathcal{A}_5=A_3-A_4,\hspace{0.5cm} (2P\cdot
x)^2\mathcal{A}_6=A_1-A_2+A_3+A_4-A_5+A_6 \nonumber
\end{eqnarray}
for the axial vector distribution amplitudes, and
\begin{eqnarray}
&&\mathcal{T}_1=T_1, \hspace{3.85cm}2P\cdot
x\mathcal{T}_2=T_1+T_2-2T_3, \nonumber\\&&
2\mathcal{T}_3=T_7,\hspace{3.68cm} 2P\cdot
x\mathcal{T}_4=T_1-T_2-2T_7, \nonumber\\&& 2P\cdot
x\mathcal{T}_5=-T_1+T_5+2T_8, \hspace{0.5cm}(2P\cdot
x)^2\mathcal{T}_6=2T_2-2T_3-2T_4+2T_5+2T_7+2T_8, \nonumber\\&& 4P
\cdot x\mathcal{T}_7=T_7-T_8, \hspace{1.90cm}(2P\cdot
x)^2\mathcal{T}_8=-T_1+T_2 +T_5-T_6+2T_7+2T_8 \nonumber
\end{eqnarray}
for the tensor distribution amplitudes. The light-cone distribution
amplitudes $F=V_i$, $A_i$, $T_i$, $S_i$, $P_i$ can be represented as
\begin{equation}
F(a_ip\cdot x)=\int \mathcal{D}x e^{-ip\cdot
x\Sigma_ix_ia_i}F(x_i)\; ,
\end{equation}
with
\begin{eqnarray}
\mathcal{D}x=dx_1dx_2dx_3\delta(x_1+x_2+x_3-1). \nonumber
\end{eqnarray}
The  distribution amplitudes are scale dependent and can be expanded
with the operators of increasing conformal spin, we write down the
explicit expressions for the $V_i$, $A_i$, $T_i$, $S_i$ and $P_i$ up
to the next-to-leading conformal spin accuracy in the appendix
\cite{BFMS,Huang04}; in the following, we will denote "the
light-cone distribution amplitudes including the next-to-leading
conformal spin" as "the $P$-wave approximation". The $V_1$, $A_1$
and $T_1$ are the leading twist-3 distribution amplitudes; the
$S_1$, $P_1$, $V_2$, $V_3$, $A_2$, $A_3$, $T_2$, $T_3$ and $T_7$ are
the twist-4 distribution amplitudes; the $S_2$, $P_2$, $V_4$, $V_5$,
$A_4$, $A_5$, $T_4$, $T_5$ and $T_8$ are the twist-5 distribution
amplitudes; while the twist-6 distribution amplitudes are the $V_6$,
$A_6$ and $T_6$. The parameters $\phi_3^0$, $\phi_6^0$, $\phi_4^0$,
$\phi_5^0$, $\xi_4^0$, $\xi_5^0$, $\psi_4^0$, $\psi_5^0$,
$\phi_3^-$, $\phi_3^+$, $\phi_4^-$, $\phi_4^+$, $\psi_4^-$,
$\psi_4^+$, $\xi_4^-$, $\xi_4^+$, $\phi_5^-$, $\phi_5^+$,
$\psi_5^-$, $\psi_5^+$, $\xi_5^-$, $\xi_5^+$, $\phi_6^-$, $\phi_6^+
$ in the light-cone distribution amplitudes $V_i$, $A_i$, $T_i$,
$S_i$, $P_i$ can be expressed in terms of eight independent matrix
elements of the local operators with the parameters $f_N$,
$\lambda_1$, $\lambda_2$, $V_1^d$, $A_1^u$, $f_d^1$, $f_d^2$ and
$f_u^1$, the three parameters $f_N$, $\lambda_1$ and $\lambda_2$ are
related to the leading order (or $S$-wave) contributions of the
conformal spin expansion,  the remaining five parameters $V_1^d$,
$A_1^u$, $f_d^1$, $f_d^2$ and $f_u^1$ are related to the
next-to-leading order (or $P$-wave) contributions of the conformal
spin expansion; the explicit expressions are given in the appendix;
for the details, one can consult Ref.\cite{BFMS}.

Taking into account the three valence  quark light-cone distribution
amplitudes up to twist-6 and performing the integration over the $x$
in the coordinate space, finally we  obtain the following results,
\begin{eqnarray}
&&z^\mu \Pi_\mu(P,q)\nonumber\\
 &=&  \!\not\!{z}\gamma_5 (P\cdot
z)^2 N(P)\left\{2\int_0^1dt_2t_2\int_0^{1-t_2}dt_1
\left[\frac{V_1-A_1+2T_1}{(q-t_2P)^2}+M^2\frac{V^u_1-A^u_1+2T^u_1}{(q-t_2P)^4}\right]\right.\nonumber\\
&&+\left.2M^2\int_0^1 d\lambda \lambda^2\int_1^\lambda dt_2
\int_0^{1-t_2}dt_1 \frac{1}{(q-\lambda P)^4}
\right.\nonumber \\
&&\left.\left[-V_1+V_4+V_5+A_1+A_4-A_5-2T_1+2T_2-2T_3+2T_5+2T_7+4T_8
\right]\right. \nonumber \\
 &&-\left.4M^2\int_0^1 d\tau \tau \int_1^\tau d\lambda \int_1^\lambda dt_2
\int_0^{1-t_2}dt_1\frac{ T_1-T_3-T_4+T_5+T_7+T_8}{(q-\tau P)^4}
\right. \nonumber \\
&&+\left.8M^2\int_0^1 d\tau \tau \int_1^\tau d\lambda \int_1^\lambda
dt_2 \int_0^{1-t_2}dt_1\frac{q^2-\tau^2M^2 }{(q-\tau
P)^4}\left[T_1-T_3-T_4+T_5+T_7+T_8\right]
\right\} \nonumber \\
&+& \!\not\!{z}\!\not\!{q}\gamma_5 (P\cdot z q\cdot z)
N(P)\left\{2M\int_0^1 d\lambda \int_1^\lambda dt_2
\int_0^{1-t_2}dt_1 \frac{1}{(q-\lambda P)^4}
\right.\nonumber \\
&&\left.\left\{V_1-V_2-V_3-A_1+A_2-A_3+2T_1-2T_3-2T_7
\right\}\right. \nonumber \\
&&+\left.8M^3\int_0^1 d\tau \tau \int_1^\tau d\lambda \int_1^\lambda
dt_2 \int_0^{1-t_2}dt_1\frac{1 }{(q-\tau
P)^6}\left[-V_1+V_2+V_3+V_4+V_5\right.\right.\nonumber\\
&&\left.\left.-V_6+A_1-A_2+A_3+A_4-A_5+A_6-2T_1+2T_3+2T_4-2T_6+2T_7+2T_8\right]
\right\} \nonumber \\
&+&\cdots ,
\end{eqnarray}
here the $V_i=V_i(t_1,t_2,1-t_1-t_2)$, $A_i=A_i(t_1,t_2,1-t_1-t_2)$
and  $T_i=T_i(t_1,t_2,1-t_1-t_2)$.

 According to the basic assumption of current-hadron duality in
the QCD sum rules approach \cite{SVZ79}, we insert  a complete
series of intermediate states satisfying the unitarity   principle
with the same quantum numbers as the current operator $\eta(0)$
 into the correlation function in
Eq.(3)  to obtain the hadronic representation. After isolating the
pole term of the lowest neutron  state, we obtain the following
result,
\begin{eqnarray}
z^\mu\Pi_\mu(P,q)&=&\frac{\!\not\!{z}P'\cdot z f_N N(P')\langle
N(P')|\bar{d}(0)\!\not\!{z} \gamma_5
u(0)|N(P)\rangle}{M^2-(q-P)^2}+\cdots
\nonumber \\
&=&\frac{P\cdot z f_N \left\{ 2P\cdot z G_A(t)\!\not\!{z}+\frac{q
\cdot z}{2M}G_P(t)\!\not\!{z}\!\not\!{q} \right\}\gamma_5N(P)}{
M^2-(q-P)^2}+\cdots \, .
\end{eqnarray}
We choose the tensor  structure $\!\not\!{z} \gamma_5 (P\cdot z)^2$
and $\!\not\!{z} \!\not\!{q} \gamma_5 (P\cdot z q\cdot z)$ to
analyze the axial form-factor $G_A(t=-Q^2)$ and induced pseudoscalar
form-factor $G_P(t=-Q^2)$ respectively.

The Borel transformation and the continuum states subtraction can be
performed by using the following substitution rules,
\begin{eqnarray}
\int dx \frac{\rho(x)}{(q-xP)^2}&=&-\int_0^1 \frac{dx}{x}
\frac{\rho(x)}{s-{P'}^2}\Rightarrow -\int_{x_0}^1 \frac{dx}{x}
\rho(x)e^{-\frac{s}{M_B^2}} , \nonumber \\
\int dx \frac{\rho(x)}{(q-xP)^4}&=&\int_0^1 \frac{dx}{x^2}
\frac{\rho(x)}{(s-{P'}^2)^2}\Rightarrow  \frac{1}{M_B^2}\int_{x_0}^1
\frac{dx}{x^2}
\rho(x)e^{-\frac{s}{M_B^2}}+\frac{\rho(x_0)e^{-\frac{s_0}{M_B^2}}}{Q^2+x_0^2
M^2} , \nonumber\\
\int dx \frac{\rho(x)}{(q-xP)^6}&=&-\int_0^1 \frac{dx}{x^3}
\frac{\rho(x)}{(s-{P'}^2)^3}\Rightarrow
-\frac{1}{2M_B^4}\int_{x_0}^1 \frac{dx}{x^3}
\rho(x)e^{-\frac{s}{M_B^2}}\nonumber \\
&&-\frac{\rho(x_0)e^{-\frac{s_0}{M_B^2}}}{2x_0(Q^2+x_0^2
M^2)}+\frac{x_0^2}{2(Q^2+x_0^2M^2)}\left[\frac{d}{dx_0}\frac{\rho(x_0)}{x_0(Q^2+x_0^2M^2)}\right] e^{-\frac{s_0}{M_B^2}}, \nonumber\\
 s&=&(1-x)M^2+\frac{(1-x)}{x}Q^2, \nonumber\\
x_0&=&\frac{\sqrt{(Q^2+s_0-M^2)^2+4M^2Q^2}-(Q^2+s_0-M^2)}{2M^2}.
\end{eqnarray}
Finally we obtain the sum rule for the axial form-factor
$G_A(t=-Q^2)$ and induced pseudoscalar form-factor $G_P(t=-Q^2)$ ,
\begin{eqnarray}
&&G_A(t)f_N e^{-\frac{M^2}{M_B^2}}\nonumber\\
&=& - \int_{x_0}^1dt_2\int_0^{1-t_2}dt_1 \exp \left\{-\frac{t_2(1-t_2)M^2+(1-t_2)Q^2}{t_2M_B^2}\right\}\left[V_1-A_1+2T_2\right]\nonumber\\
&+&\frac{M^2}{M_B^2} \int_{x_0}^1\frac{dt_2}{t_2}\int_0^{1-t_2}dt_1 \exp \left\{-\frac{t_2(1-t_2)M^2+(1-t_2)Q^2}{t_2M_B^2}\right\}\left[V^u_1-A^u_1+2T^u_2\right]\nonumber\\
&+&\frac{x_0M^2}{Q^2+x_0^2M^2} \int_0^{1-x_0}dt_1 \exp \left\{-\frac{s_0}{M_B^2}\right\}\left[V^u_1-A^u_1+2T^u_2\right]\nonumber\\
&+& \frac{M^2}{M_B^2} \int_{x_0}^1d\lambda \int_1^\lambda dt_2\int_0^{1-t_2}dt_1 \exp \left\{-\frac{\lambda(1-\lambda)M^2+(1-\lambda)Q^2}{\lambda M_B^2}\right\}\nonumber \\
&&\left[-V_1+V_4+V_5+A_1+A_4-A_5-2T_1+2T_2-2T_3+2T_5+2T_7+4T_8\right]\nonumber\\
&+& \frac{x_0^2M^2}{Q^2+x_0^2M^2}  \int_1^{x_0} dt_2\int_0^{1-t_2}dt_1 \exp \left\{-\frac{s_0}{M_B^2}\right\}\nonumber \\
&&\left[-V_1+V_4+V_5+A_1+A_4-A_5-2T_1+2T_2-2T_3+2T_5+2T_7+4T_8\right]\nonumber\\
  &-& \frac{2M^2}{M_B^2} \int_{x_0}^1\frac{\tau}{\tau}\int_1^\tau d\lambda \int_1^\lambda dt_2\int_0^{1-t_2}dt_1
  \exp \left\{-\frac{\tau(1-\tau)M^2+(1-\tau)Q^2}{\tau M_B^2}\right\}\nonumber \\
&&\left[T_1-T_3-T_4+T_5+T_7+T_8\right]\nonumber\\
&-& \frac{2x_0M^2}{Q^2+x_0^2M^2} \int_1^{x_0} d\lambda
\int_1^\lambda dt_2\int_0^{1-t_2}dt_1
  \exp \left\{-\frac{s_0}{ M_B^2}\right\} \left[T_1-T_3-T_4+T_5+T_7+T_8\right]\nonumber\\
  &+& \frac{2M^2}{M_B^4} \int_{x_0}^1\frac{\tau}{\tau^2}\int_1^\tau d\lambda \int_1^\lambda dt_2\int_0^{1-t_2}dt_1
  \exp \left\{-\frac{\tau(1-\tau)M^2+(1-\tau)Q^2}{\tau M_B^2}\right\}\nonumber \\
&&\left[Q^2+\tau^2M^2\right]\left[T_1-T_3-T_4+T_5+T_7+T_8\right]\nonumber\\
&+& \frac{2M^2}{M_B^2} \int_1^{x_0} d\lambda \int_1^\lambda
dt_2\int_0^{1-t_2}dt_1
  \exp \left\{-\frac{s_0}{ M_B^2}\right\}\left[T_1-T_3-T_4+T_5+T_7+T_8\right]\nonumber\\
  &-& \frac{2x_0^2M^2}{Q^2+x_0^2M^2}  \int_1^{x_0}
dt_2\int_0^{1-t_2}dt_1
  \exp \left\{-\frac{s_0}{
  M_B^2}\right\}\left[T_1-T_3-T_4+T_5+T_7+T_8\right].
\end{eqnarray}

\begin{eqnarray}
&&G_P(t)f_N e^{-\frac{M^2}{M_B^2}}\nonumber\\
&=&  \frac{4M^2}{M_B^2} \int_{x_0}^1 \frac{d\lambda}{\lambda^2} \int_1^\lambda dt_2\int_0^{1-t_2}dt_1 \exp \left\{-\frac{\lambda(1-\lambda)M^2+(1-\lambda)Q^2}{\lambda M_B^2}\right\}\nonumber \\
&&\left[V_1-V_2-V_3-A_1+A_2-A_3+2T_1-2T_3-2T_7\right]\nonumber\\
&+& \frac{4M^2}{Q^2+x_0^2M^2}  \int_1^{x_0} dt_2\int_0^{1-t_2}dt_1 \exp \left\{-\frac{s_0}{M_B^2}\right\}\nonumber \\
&&\left[V_1-V_2-V_3-A_1+A_2-A_3+2T_1-2T_3-2T_7\right]\nonumber\\
  &-& \frac{8M^4}{M_B^4} \int_{x_0}^1\frac{\tau}{\tau^2}\int_1^\tau d\lambda \int_1^\lambda dt_2\int_0^{1-t_2}dt_1
  \exp \left\{-\frac{\tau(1-\tau)M^2+(1-\tau)Q^2}{\tau M_B^2}\right\}\nonumber \\
&&\left[-V_1+V_2+V_3+V_4+V_5-V_6+A_1-A_2+A_3+A_4-A_5+A_6\right.\nonumber\\
&&\left.-2T_1+2T_3+2T_4-2T_6+2T_7+2T_8\right]\nonumber\\
&-& \frac{8M^4}{(Q^2+x_0^2M^2)M_B^2} \int_1^{x_0} d\lambda
\int_1^\lambda dt_2\int_0^{1-t_2}dt_1
  \exp \left\{-\frac{s_0}{ M_B^2}\right\}\left[-V_1+V_2+V_3+V_4+V_5\right.\nonumber\\
&&\left.-V_6+A_1-A_2+A_3+A_4-A_5+A_6-2T_1+2T_3+2T_4-2T_6+2T_7+2T_8\right]\nonumber\\
  &-& \frac{8x_0^2M^4}{Q^2+x_0^2M^2} \left[\frac{d}{dx_0} \frac{1}{Q^2+x_0^2}\int_1^{x_0} d\lambda
\int_1^\lambda dt_2\int_0^{1-t_2}dt_1\right]
  \exp \left\{-\frac{s_0}{ M_B^2}\right\}\left[-V_1+V_2+V_3\right.\nonumber\\
&&\left.+V_4+V_5-V_6+A_1-A_2+A_3+A_4-A_5+A_6-2T_1+2T_3+2T_4-2T_6+2T_7+2T_8\right]. \nonumber\\
\end{eqnarray}
\section{Numerical results and discussions}
The input parameters have to be specified  before the numerical
analysis. We choose the suitable range  for the Borel parameter
$M_B$, $1.5GeV^2<M_B^2<2.5GeV^2$. In this range, the Borel parameter
$M_B$ is small enough  to warrant the higher mass resonances and
 continuum states are  suppressed sufficiently, on the other hand,  it is
  large enough to warrant the convergence of the
light-cone expansion with increasing twists in the perturbative QCD
calculation \cite{Ioffe84,Ioffe81}. The numerical results indicate
that in this range the  form-factors $G_A(t=-Q^2)$ and $G_P(t=-Q^2)$
are almost independent on the Borel parameter $M_B$, which we can
see from the Fig.1 and Fig.2 respectively for the central values of
the eight input parameters $f_N$, $\lambda_1$, $\lambda_2$, $V_1^d$,
$A_1^u$, $f_d^1$, $f_d^2$ and $f_u^1$.
\begin{figure}
 \centering
 \includegraphics[totalheight=7cm]{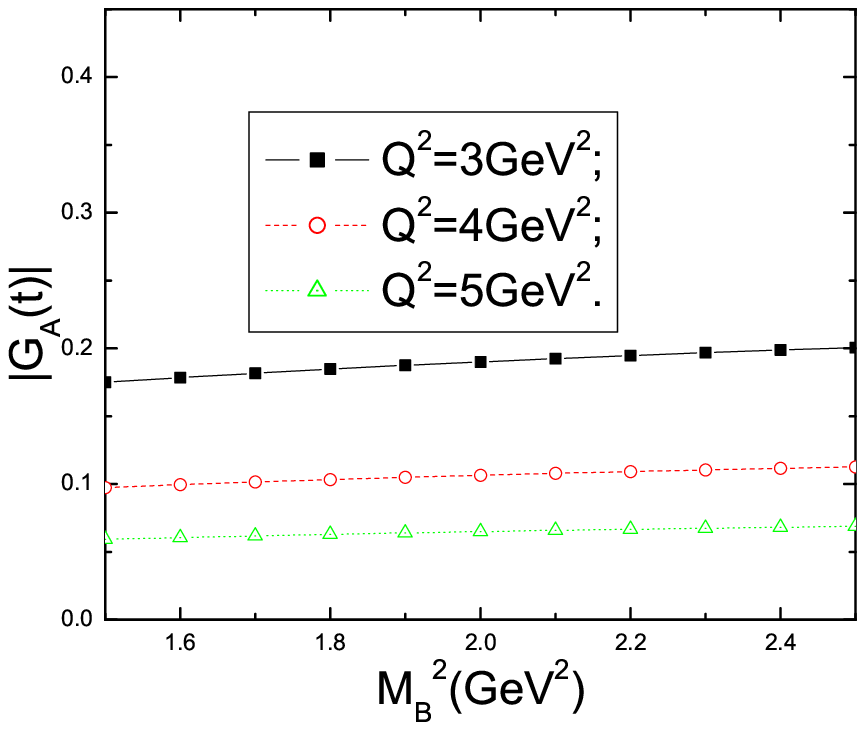}
 \caption{The  axial form-factor $G_A(t)$ with the parameter $M_B$ for $s_0=2.25GeV^2$. }
\end{figure}
\begin{figure}
 \centering
 \includegraphics[totalheight=7cm]{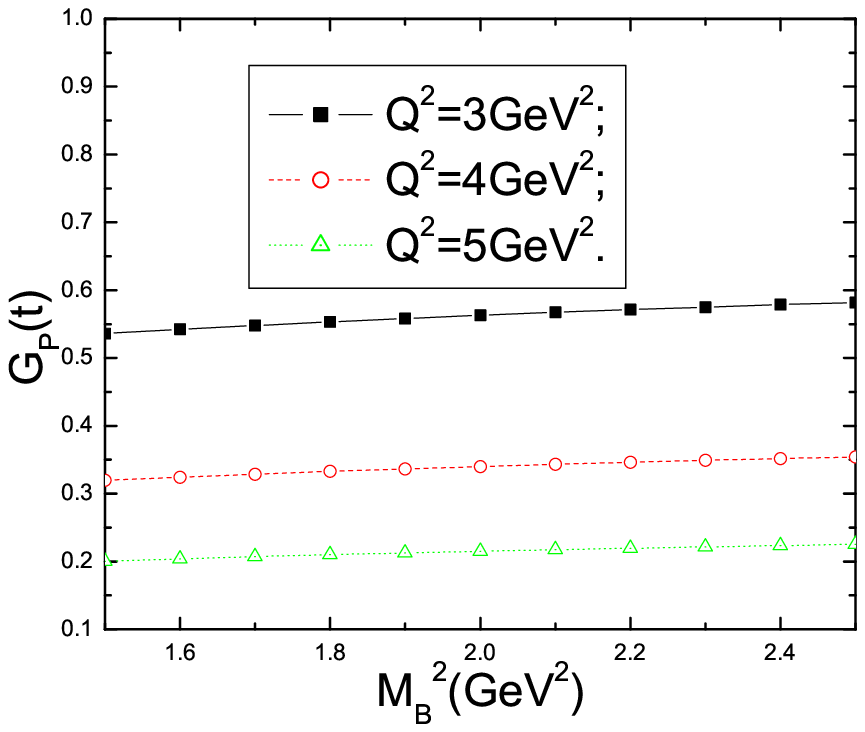}
 \caption{The  induced pseudoscalar form-factor $G_P(t)$ with the  parameter $M_B$ for $s_0=2.25GeV^2$. }
\end{figure}
 For simplicity, we choose
the standard  values for the  threshold parameter $s_0$,
$s_0=2.25GeV^2$,  to
 subtract the contributions from the higher resonances and
 continuum states i.e. we restrict the range of integral  to the energy region
 below the Roper resonance ($N(1440)$);  furthermore, it is large enough to
take into account  all contributions from the neutron.   For
$Q^2=(2-9)GeV^2$, $x\geq x_0=0.5-0.8$, the average value $\langle
x\rangle=0.75-0.90$,
   with the intermediate and
large space-like momentum $Q^2$, the end-point (soft) contributions
(or the Feynman mechanism) are dominant, it is consistent with the
growing consensus that the onset of the perturbative QCD region in
exclusive processes is postponed to very large energy scales.
 The parameters in the light-cone distribution amplitudes  $ \phi_3^0 $,
  $\phi_6^0$, $\phi_4^0$, $\phi_5^0$,  $\xi_4^0$, $\xi_5^0$, $\psi_4^0$,
$\psi_5^0 $,  $ \phi_3^-$, $\phi_3^+$, $\phi_4^-$, $\phi_4^+$,
$\psi_4^-$, $\psi_4^+$, $\xi_4^-$, $\xi_4^+$, $\phi_5^-$,
$\phi_5^+$, $\psi_5^-$, $\psi_5^+$, $\xi_5^-$ ,$ \xi_5^+$,
$\phi_6^-$, $\phi_6^+ $ are scale dependent and can be calculated
with the corresponding QCD sum rules. They are functions of eight
independent parameters $f_N$, $\lambda_1$, $\lambda_2$, $V_1^d$,
$A_1^u$, $f^d_1$, $f^d_2$ and $f^u_1$,  the three parameters $f_N$,
$\lambda_1$ and $\lambda_2$ are related to the leading order (or
$S$-wave) contributions in the conformal spin expansion,  the
remaining five parameters $V_1^d$, $A_1^u$, $f^d_1$, $f^d_2$ and
$f^u_1$ are related to the next-to-leading order (or $P$-wave)
contributions in the conformal spin expansion; the explicit
expressions are presented in the appendix, for detailed and
systematic studies about this subject, one can consult
Ref.\cite{BFMS}.  Here we take the values at the energy scale
$\mu=1GeV$ and neglect the  evolution with the energy scale $\mu$
for simplicity, the values for the eight independent parameters are
taken as $f_N=(5.3\pm 0.5)\times 10^{-3} GeV^2$, $\lambda_1=-(2.7\pm
0.9)\times 10^{-2}GeV^2$, $\lambda_2=(5.1\pm 1.9)\times
10^{-2}GeV^2$, $V_1^d=0.23\pm 0.03$, $A_1^u=0.38\pm 0.15$,
$f_1^d=0.6\pm 0.2$, $f_2^d=0.15\pm 0.06$ and $f_1^u=0.22\pm 0.15$.
In estimating those coefficients with the QCD sum rules, only the
first few moments are taken into account, the values are not very
accurate. In the limit $Q^2\rightarrow \infty$, the five parameters
related to the  light-cone distribution amplitudes with the $P$-wave
conformal spin take the asymptotic values $f^d_1=\frac{3}{10}$,
$f^d_2=\frac{4}{15}$, $f^u_1=\frac{1}{10}$, $A^u_1=0$ and
$V^d_1=\frac{1}{3}$.

We perform the operator product expansion in the light-cone with
large $Q^2$ and $P'^2$, the form-factors $G_A(t=-Q^2)$ and
$G_P(t=-Q^2)$ make sense at the regions, for example,  $Q^2>2GeV^2$,
with low momentum transfers, the operator product expansion is
questionable. In numerical analysis, we observe that  the axial
form-factor $G_A(t=-Q^2)$ is sensitive to the two parameters
$\lambda_1$ and $f^d_1$, small variations of the two parameters can
lead to relatively large changes for the values, the induced
pseudoscalar form-factor $G_P(t=-Q^2)$ is sensitive to the four
parameters $f_N$, $\lambda_1$, $f^d_1$ and $f^u_1$, small variations
of those parameters, especially the $\lambda_1$ and $f^d_1$, can
lead to large changes for the values, which are shown in the Fig.3,
Fig.4, Fig.5, Fig.6, Fig.7 and Fig.8, respectively. The   large
uncertainties can impair the predictive ability of the sum rules,
 the parameters $\lambda_1$, $f^d_1$, $f_N$ and $f^u_1$ should
be refined to make robust predictions, in Ref.\cite{WangWY}, we
observe that the scalar-form factor of the nucleon is sensitive to
the four parameters $\lambda_1$, $f^d_1$, $f^d_2$ and $f^u_1$, so
refining the three parameters $\lambda_1$, $f^d_1$ and $f^u_1$ is of
great importance. The final numerical values for the axial
form-factor $G_A(t=-Q^2)$ and induced pseudoscalar form-factor
$G_P(t=-Q^2)$ at the intermediate and large space-like momentum
regions, $2GeV^2<Q^2<9GeV^2$, are plotted in the Fig.9 and Fig.10
respectively.

From those figures, we can see that the central values of the axial
form-factor $G_A(t=-Q^2)$ lie above the results of the double-pole
fitted formulation from the neutrino scattering experiments
\cite{Review2002},
\begin{eqnarray}
G_A(t)=\frac{g_A}{(1-t/M_A^2)^2}\, ,
\end{eqnarray}
here we take the values $g_A=1.2673$, $M_A=1.026$, and neglect the
uncertainties for simplicity; at the region $Q^2>4.0GeV^2$, the
values of the double-pole fitted formulation lie between the up and
down limits, our results can make both qualitative and quantitative
predictions. Furthermore, our results are compatible with the
calculation of lattice QCD \cite{DLL96} and chiral quark models
\cite{ChiralAxial}. For the induced pseudoscalar form-factor
$G_P(t=-Q^2)$, the uncertainties are very large and the values make
sense only qualitatively, not quantitatively, our results are
compatible with the calculation from the Bethe-Salpeter equation
\cite{BSall}.

In the limit $Q^2\rightarrow\infty$, we present the numerical values
 for the axial form-factor $G_A(t=-Q^2)$ and the induced pseudoscalar form-factor
$G_P(t=-Q^2)$ with the  asymptotic light-cone distribution
amplitudes in the Fig.11 and Fig.12, respectively. From the Fig.11,
 we can see that for the axial form-factor $G_A(t=-Q^2)$, the
values with the asymptotic light-cone distribution amplitudes lie
above the corresponding ones with the light-cone distribution
amplitudes in the $P$-wave approximation, at $Q^2
> 10GeV^2$, the two curves approach the  values of the double-pole
fitted  formulation $G_A(t=-Q^2)\sim \frac{1}{Q^4}$, which is
expected from the naive power counting rules. From the Fig.12, we
can see that for the induced pseudoscalar form-factor $G_P(t=-Q^2)$,
the values with the asymptotic light-cone distribution amplitudes
have negative sign comparing  with the corresponding ones with the
light-cone distribution amplitudes in the $P$-wave approximation, at
$Q^2 > 10GeV^2$, the two curves approach the same values.  The large
difference between the values from the asymptotic light-cone
distribution amplitudes and the $P$-wave approximated light-cone
distribution amplitudes again indicate  the importance of the
contributions from the $P$-wave conformal spin at the intermediate
and large momentum transfers $2GeV^2<Q^2<10GeV^2$, to make robust
predictions, we have to refine the five parameters.

The consistent and complete LCSR analysis should take into account
the contributions from the perturbative $\alpha_s$ corrections, the
distribution amplitudes with additional valence gluons and
quark-antiquark pairs, and improve the parameters which enter in the
LCSRs.
\begin{figure}
 \centering
 \includegraphics[totalheight=7cm]{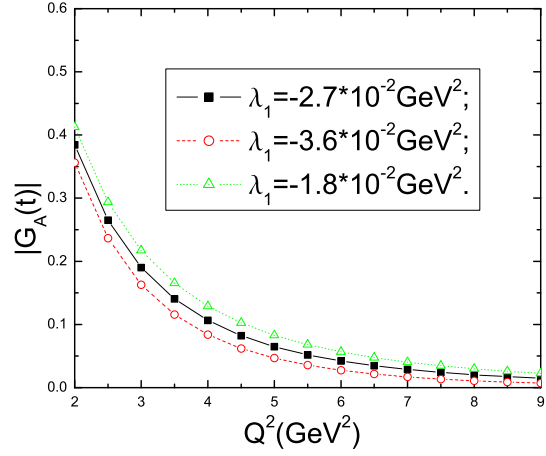}
 \caption{The axial form-factor $G_A(t)$ with the parameters $M_B^2=2.0GeV^2$ and $\lambda_1$.}
\end{figure}

\begin{figure}
 \centering
 \includegraphics[totalheight=7cm]{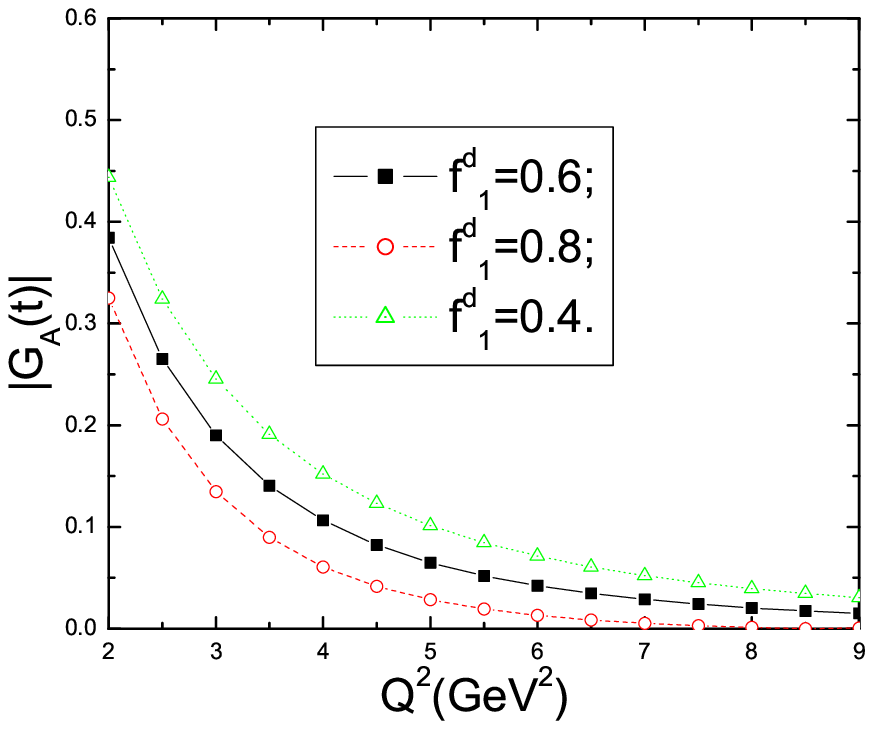}
 \caption{The axial form-factor $G_A(t)$ with the parameters $M_B^2=2.0GeV^2$ and $f^d_1$.}
\end{figure}

\begin{figure}
 \centering
 \includegraphics[totalheight=7cm]{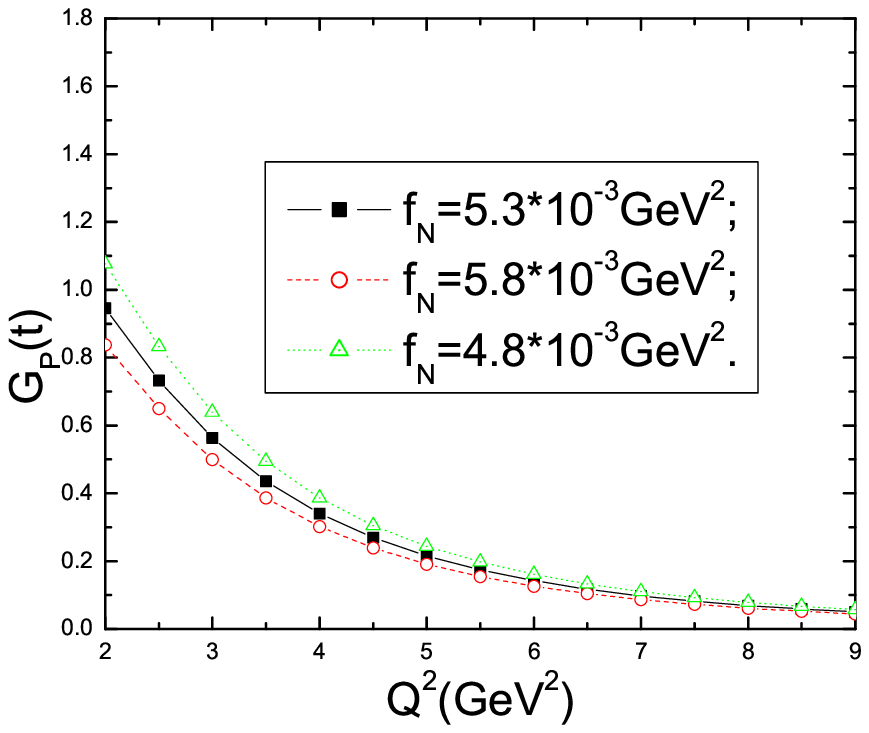}
 \caption{The induced pseudoscalar form-factor $G_P(t)$ with the parameters $M_B^2=2.0GeV^2$ and $f_N$.}
\end{figure}

\begin{figure}
 \centering
 \includegraphics[totalheight=7cm]{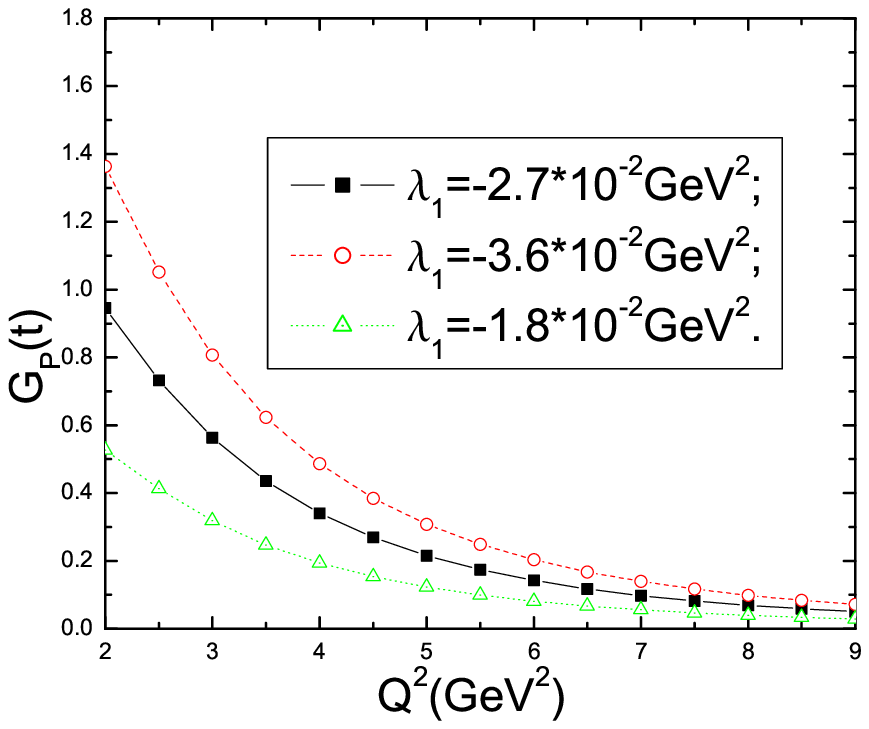}
 \caption{The induced pseudoscalar form-factor $G_P(t)$ with the parameters $M_B^2=2.0GeV^2$ and $\lambda_1$.}
\end{figure}

\begin{figure}
 \centering
 \includegraphics[totalheight=7cm]{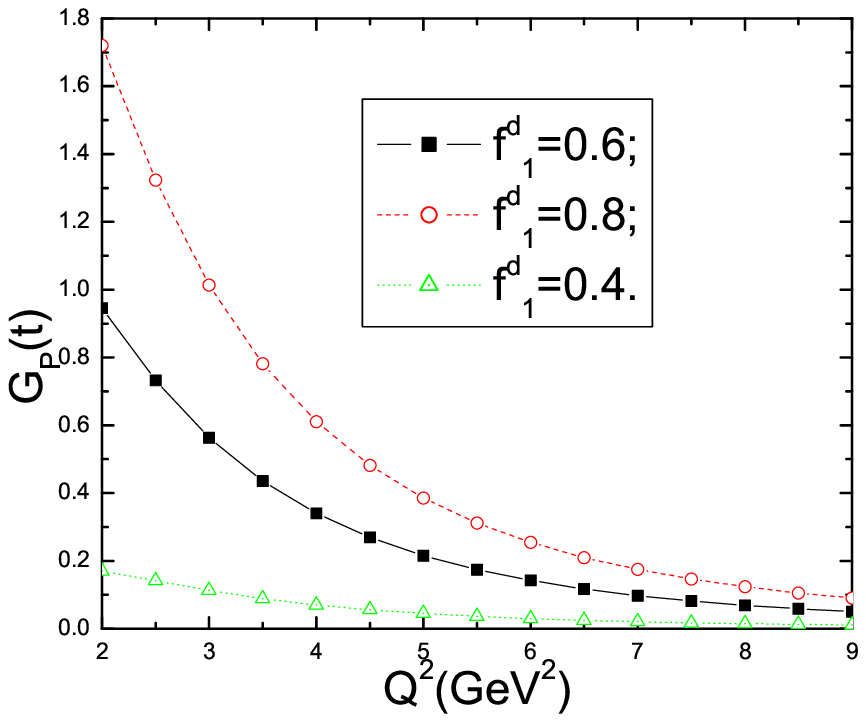}
 \caption{The induced pseudoscalar form-factor $G_P(t)$ with the parameters $M_B^2=2.0GeV^2$ and $f^d_1$.}
\end{figure}

\begin{figure}
 \centering
 \includegraphics[totalheight=7cm]{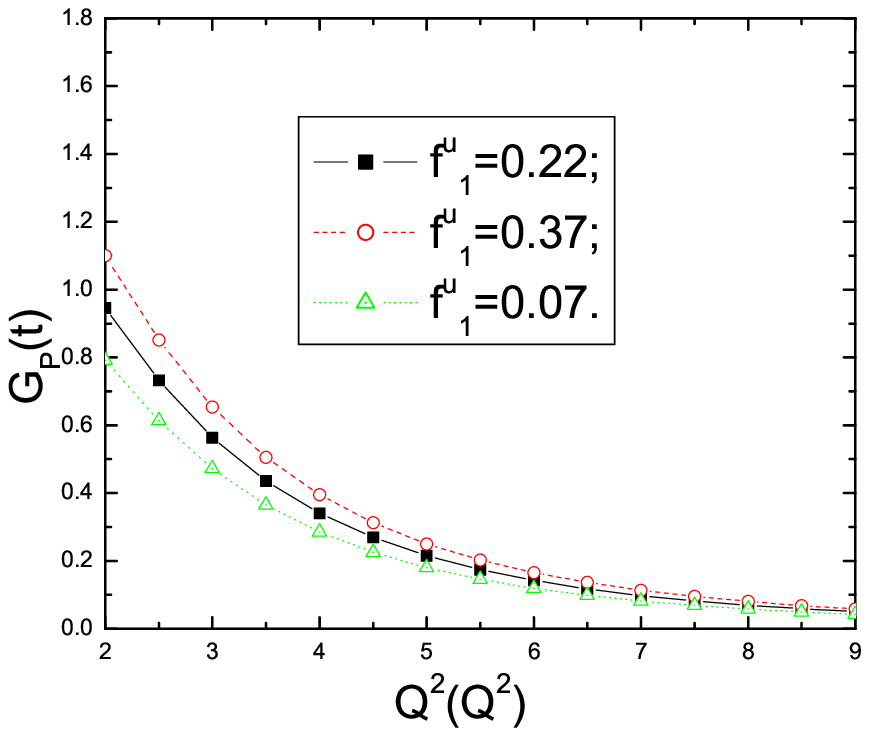}
 \caption{The induced pseudoscalar form-factor $G_P(t)$ with the parameters $M_B^2=2.0GeV^2$ and $f^u_1$.}
\end{figure}

\begin{figure}
 \centering
 \includegraphics[totalheight=7cm]{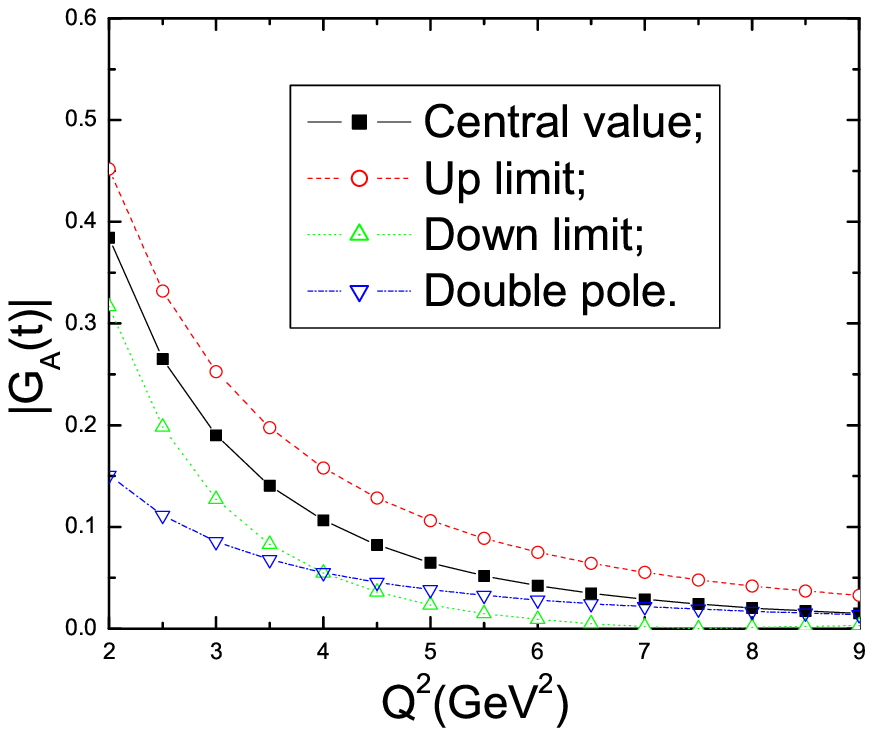}
 \caption{The axial form-factor $G_A(t)$ with the parameter $M_B^2=2.0GeV^2$.}
\end{figure}

\begin{figure}
 \centering
 \includegraphics[totalheight=7cm]{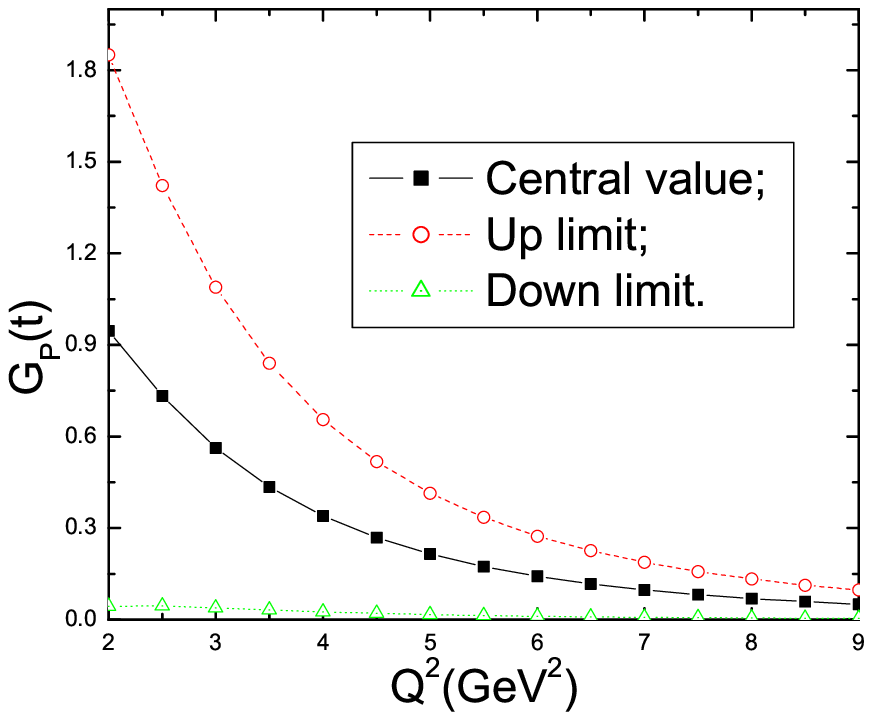}
 \caption{The  induced pseudoscalar form-factor $G_P(t)$ with the parameter $M_B^2=2.0GeV^2$.}
\end{figure}

\begin{figure}
 \centering
 \includegraphics[totalheight=7cm]{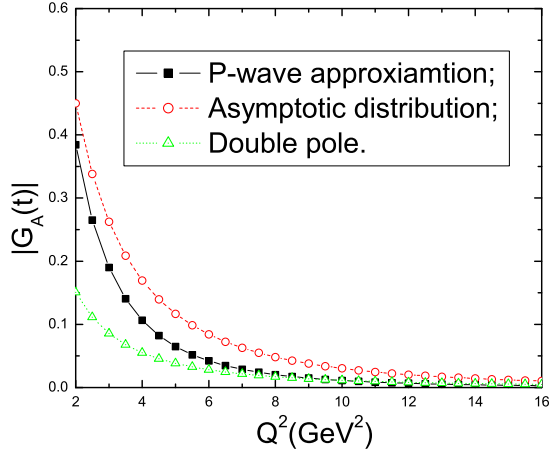}
 \caption{The axial form-factor $G_A(t)$ with the parameters $f_N=5.3\times 10^{-3} GeV^2$, $\lambda_1=-2.7\times 10^{-2}GeV^2$,
 $\lambda_2=5.1\times
10^{-2}GeV^2$ and $M_B^2=2.0GeV^2$.}
\end{figure}

\begin{figure}
 \centering
 \includegraphics[totalheight=7cm]{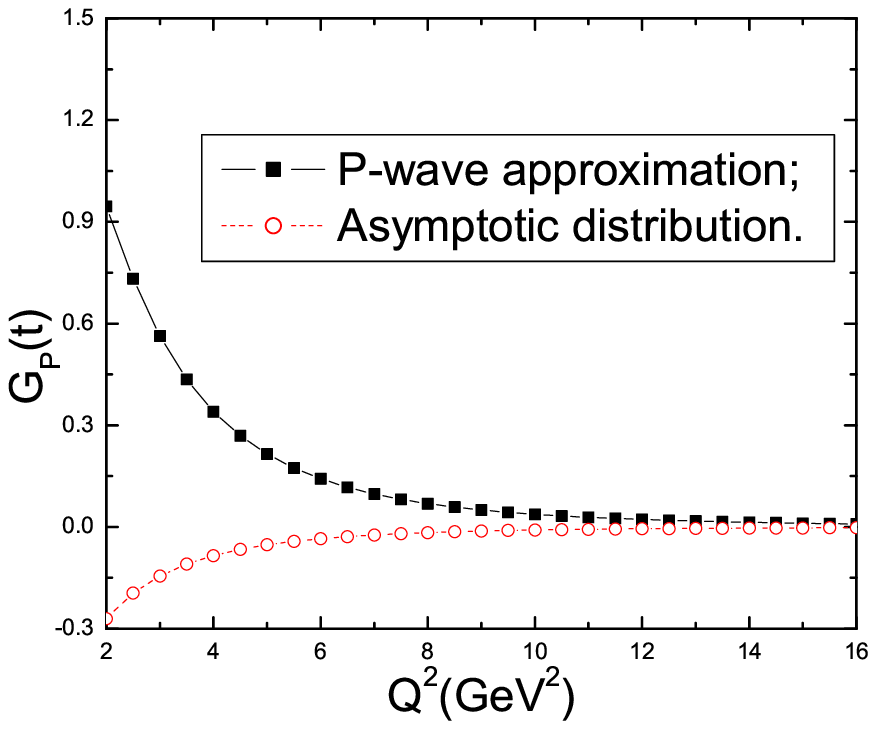}
 \caption{The  $G_P(t)$ with the parameters $f_N=5.3\times 10^{-3} GeV^2$, $\lambda_1=-2.7\times 10^{-2}GeV^2$,
 $\lambda_2=5.1\times
10^{-2}GeV^2$ and $M_B^2=2.0GeV^2$.}
\end{figure}

\section{Conclusion }

In this work, we  calculate the
  axial and induced pseudoscalar form-factors $G_A(t=-Q^2)$ and $G_P(t=-Q^2)$ of the nucleons in the
  framework of the LCSR approach up
 to twist-6 three valence quark light-cone distribution amplitudes.
  The  form-factors $G_A(t=-Q^2)$ and $G_P(t=-Q^2)$ at intermediate and
large momentum transfers with $Q^2> 2 GeV^2$ have significant
contributions from the end-point (soft) terms. The axial form-factor
$G_A(t=-Q^2)$ is sensitive to the two parameters $\lambda_1$ and
$f^d_1$, small variations of the two parameters can lead to
relatively large changes for the values; the induced pseudoscalar
form-factor $G_P(t=-Q^2)$ is sensitive to the four parameters $f_N$,
$\lambda_1$, $f^d_1$ and $f^u_1$, small variations of those
parameters, especially the $\lambda_1$ and $f^d_1$, can lead to
large changes for the values. The   large uncertainties can impair
the predictive ability of the sum rules,  the parameters
$\lambda_1$, $f^d_1$, $f_N$ and $f^u_1$ should be refined to make
robust predictions. The numerical values for the  $G_A(t=-Q^2)$  are
 compatible with the experimental data and  theoretical calculations, for
 example, chiral quark model and lattice QCD. In the limit $Q^2\rightarrow\infty$,
  the values for  the axial form-factor $G_A(t=-Q^2)$ with both the asymptotic
light-cone distribution amplitudes and the light-cone distribution
amplitudes in the $P$-wave approximation approach the results of the
double-pole fitted  formulation $G_A(t=-Q^2)\sim \frac{1}{Q^4}$.
 The numerical results for the induced pseudoscalar form-factor
 $G_P(t=-Q^2)$ are compatible with the calculation from the
 Bethe-Salpeter equation, in the limit $Q^2\rightarrow\infty$,
  the values of the  $G_A(t=-Q^2)$ with both the asymptotic
light-cone distribution amplitudes and the light-cone distribution
amplitudes in the $P$-wave approximation approach the same values.
 The consistent and complete LCSR analysis should take into account
the contributions from the perturbative $\alpha_s$ corrections, the
distribution amplitudes with additional valence gluons and
quark-antiquark pairs, and improve the parameters which enter in the
LCSRs.

\section*{Acknowledgment}
This  work is supported by National Natural Science Foundation,
Grant Number 10405009,  and Key Program Foundation of NCEPU. The
authors are indebted to Dr. J.He (IHEP), Dr. X.B.Huang (PKU) and Dr.
L.Li (GSCAS) for numerous help, without them, the work would not be
finished.
\appendix
\section*{Appendix}
 \begin{eqnarray}
V_1(x_i,\mu)&=&120x_1x_2x_3[\phi_3^0(\mu)+\phi_3^+(\mu)(1-3x_3)],\nonumber\\
V_2(x_i,\mu)&=&24x_1x_2[\phi_4^0(\mu)+\phi_3^+(\mu)(1-5x_3)],\nonumber\\
V_3(x_i,\mu)&=&12x_3\{\psi_4^0(\mu)(1-x_3)+\psi_4^-(\mu)[x_1^2+x_2^2-x_3(1-x_3)]
\nonumber\\&&+\psi_4^+(\mu)(1-x_3-10x_1x_2)\},\nonumber\\
V_4(x_i,\mu)&=&3\{\psi_5^0(\mu)(1-x_3)+\psi_5^-(\mu)[2x_1x_2-x_3(1-x_3)]
\nonumber\\&&+\psi_5^+(\mu)[1-x_3-2(x_1^2+x_2^2)]\},\nonumber\\
V_5(x_i,\mu)&=&6x_3[\phi_5^0(\mu)+\phi_5^+(\mu)(1-2x_3)],\nonumber\\
V_6(x_i,\mu)&=&2[\phi_6^0(\mu)+\phi_6^+(\mu)(1-3x_3)].\nonumber
\end{eqnarray}
\begin{eqnarray}
A_1(x_i,\mu)&=&120x_1x_2x_3\phi_3^-(\mu)(x_2-x_1),\nonumber\\
A_2(x_i,\mu)&=&24x_1x_2\phi_4^-(\mu)(x_2-x_1),\nonumber\\
A_3(x_i,\mu)&=&12x_3(x_2-x_1)\{(\psi_4^0(\mu)+\psi_4^+(\mu))+\psi_4^-(\mu)(1-2x_3)
\},\nonumber\\
A_4(x_i,\mu)&=&3(x_2-x_1)\{-\psi_5^0(\mu)+\psi_5^-(\mu)x_3
+\psi_5^+(\mu)(1-2x_3)\},\nonumber\\
A_5(x_i,\mu)&=&6x_3(x_2-x_1)\phi_5^-(\mu)\nonumber\\
A_6(x_i,\mu)&=&2(x_2-x_1)\phi_6^-(\mu).\nonumber
\end{eqnarray}
\begin{eqnarray}
T_1(x_i,\mu)&=&120x_1x_2x_3[\phi_3^0(\mu)+\frac{1}{2}(\phi_3^--\phi_3^+)(\mu)(1-3x_3)
],\nonumber\\
T_2(x_i,\mu)&=&24x_1x_2[\xi_4^0(\mu)+\xi_4^+(\mu)(1-5x_3)],\nonumber\\
T_3(x_i,\mu)&=&6x_3\{(\xi_4^0+\phi_4^0+\psi_4^0)(\mu)(1-x_3)+
(\xi_4^-+\phi_4^--\psi_4^-)(\mu)[x_1^2+x_2^2-x_3(1-x_3)]
\nonumber\\
&&+(\xi_4^++\phi_4^++\psi_4^+)(\mu)(1-x_3-10x_1x_2)\},\nonumber\\
T_4(x_i,\mu)&=&\frac{3}{2}\{(\xi_5^0+\phi_5^0+\psi_5^0)(\mu)(1-x_3)+
(\xi_5^-+\phi_5^--\psi_5^-)(\mu)[2x_1x_2-x_3(1-x_3)]
\nonumber\\
&&+(\xi_5^++\phi_5^++\psi_5^+)(\mu)(1-x_3-2(x_1^2+x_2^2))\},\nonumber\\
T_5(x_i,\mu)&=&6x_3[\xi_5^0(\mu)+\xi_5^+(\mu)(1-2x_3)],\nonumber\\
T_6(x_i,\mu)&=&2[\phi_6^0(\mu)+\frac{1}{2}(\phi_6^--\phi_6^+)(\mu)(1-3x_3)],
\nonumber \\
T_7(x_i,\mu)&=&6x_3\{(-\xi_4^0+\phi_4^0+\psi_4^0)(\mu)(1-x_3)+
(-\xi_4^-+\phi_4^--\psi_4^-)(\mu)[x_1^2+x_2^2-x_3(1-x_3)]
\nonumber\\
&&+(-\xi_4^++\phi_4^++\psi_4^+)(\mu)(1-x_3-10x_1x_2)\},\nonumber\\
T_8(x_i,\mu)&=&\frac{3}{2}\{(-\xi_5^0+\phi_5^0+\psi_5^0)(\mu)(1-x_3)+
(-\xi_5^-+\phi_5^--\psi_5^-)(\mu)[2x_1x_2-x_3(1-x_3)]
\nonumber\\
&&+(-\xi_5^++\phi_5^++\psi_5^+)(\mu)(1-x_3-2(x_1^2+x_2^2))\}.\nonumber
\end{eqnarray}
\begin{eqnarray}
S_1(x_i,\mu) &=& 6 x_3 (x_2-x_1) \left[ (\xi_4^0 + \phi_4^0 +
\psi_4^0 + \xi_4^+ + \phi_4^+ + \psi_4^+)(\mu) + (\xi_4^- + \phi_4^-
- \psi_4^-)(\mu)(1-2 x_3) \right] \,,
\nonumber \\
S_2(x_i,\mu) &=& \frac{3}{2} (x_2 -x_1) \left[- \left(\psi_5^0 +
\phi_5^0 + \xi_5^0\right)(\mu) + \left(\xi_5^- + \phi_5^- - \psi_5^0
\right)(\mu) x_3 \right. \nonumber \\
 && \left.+\left(\xi_5^+ + \phi_5^+ + \psi_5^0 \right)(\mu) (1- 2
x_3)\right]\,,
\nonumber \\
P_1(x_i,\mu) &=& 6 x_3 (x_2-x_1) \left[ (\xi_4^0 - \phi_4^0 -
\psi_4^0 + \xi_4^+ - \phi_4^+ - \psi_4^+)(\mu) + (\xi_4^- - \phi_4^-
+ \psi_4^-)(\mu)(1-2 x_3) \right] \, ,
\nonumber \\
P_2(x_i,\mu) &=& \frac32 (x_2 -x_1) \left[\left(\psi_5^0 + \psi_5^0
- \xi_5^0\right)(\mu) + \left(\xi_5^- - \phi_5^- + \psi_5^0
\right)(\mu) x_3 \right. \nonumber\\
&& \left. + \left(\xi_5^+ - \phi_5^+ - \psi_5^0 \right)(\mu) (1- 2
x_3)\right]\, . \nonumber
\end{eqnarray}
\begin{eqnarray}
 \mathcal{V}_1^{d}(x_3) &=&\frac{x_3^2}{24}(\lambda_1
C_\lambda^d+f_N C_f^d),\nonumber\\
\mathcal{V}_1^{u}(x_2)&=&\frac{x_2^2}{24}(\lambda_1 C_\lambda^u+f_N
C_f^u), \nonumber \\
C_\lambda^d&=& -(1-x_3)[11+131\,x_3-169x_3^2+63x_3^3-30\,f_1^d
\,(3+11x_3-17x_3^2+7x_3^3)]\nonumber \\
&&-12\,(3-10\,f_1^d)\,\ln x_3,\nonumber\\
C_f^d&=& -( 1 - x_3 )
\,[1441+505x_3-3371x_3^2+3405x_3^3-1104x_3^4-24V_1^d\nonumber\\
&&(207-3x_3-368x_3^2+412x_3^3-138x_3^4)]  - 12(73-220\,V_1^d)\,\ln
x_3,\nonumber\\
 C_\lambda^u &=&
-(1-x_2)^3[13-20f_1^d+3x_2+10f_1^u(1-3x_2)],\nonumber\\
 C_f^u&=&(1-x_2)^3[113+495x_2-552x_2^2+10A_1^u(-1+3x_2)\nonumber\\
 &&+2V_1^d(113-951x_2+828x_2^2)].
 \nonumber
\end{eqnarray}
\begin{eqnarray}
\mathcal{A}_1^{d}(x_3)&=&0,\nonumber\\
\mathcal{A}_1^{u}(x_2)&=&\frac{x_2^2}{24}( 1- x_2 )^3(\lambda_1
D_\lambda^u+f_N D_f^u), \nonumber \\
 D_\lambda^u&=&29-45x_2-10f_1^u(7-9x_2)-20f_1^d(5-6x_2)
,\nonumber\\
D_f^u&=&11+45x_2+10V_1^d(1-30x_2)-2A_1^u(113-951x_2+828x_2^2).
\nonumber
\end{eqnarray}
\begin{eqnarray}
\phi_3^0 = \phi_6^0 = f_N \,,\hspace{0.3cm} &\qquad& \phi_4^0 =
\phi_5^0 = \frac{1}{2} \left(\lambda_1 + f_N\right) \,,
\nonumber \\
\xi_4^0 = \xi_5^0 = \frac{1}{6} \lambda_2\,, &\qquad& \psi_4^0  =
\psi_5^0 = \frac{1}{2}\left(f_N - \lambda_1 \right)  \,. \nonumber
\end{eqnarray}
\begin{eqnarray}
\tilde\phi_3^- &=& \frac{21}{2} A_1^u,\nonumber\\
\tilde\phi_3^+ &=& \frac{7}{2} (1 - 3 V_1^d),\nonumber\\
\phi_4^- &=& \frac{5}{4} \left(\lambda_1(1- 2 f_1^d -4 f_1^u) + f_N(
2 A_1^u - 1)\right) \,,
\nonumber \\
\phi_4^+ &=& \frac{1}{4} \left( \lambda_1(3- 10 f_1^d) - f_N( 10
V_1^d - 3)\right)\,,
\nonumber \\
\psi_4^- &=& - \frac{5}{4} \left(\lambda_1(2- 7 f_1^d + f_1^u) +
f_N(A_1^u + 3 V_1^d - 2)\right) \,,
\nonumber \\
\psi_4^+ &=& - \frac{1}{4} \left(\lambda_1 (- 2 + 5 f_1^d + 5 f_1^u)
+ f_N( 2 + 5 A_1^u - 5 V_1^d)\right)\,,
\nonumber \\
\xi_4^- &=& \frac{5}{16} \lambda_2(4- 15 f_2^d)\,,
\nonumber \\
\xi_4^+ &=& \frac{1}{16} \lambda_2 (4- 15 f_2^d)\,,\nonumber
\end{eqnarray}
\begin{eqnarray}
\phi_5^- &=& \frac{5}{3} \left(\lambda_1(f_1^d - f_1^u) + f_N( 2
A_1^u - 1)\right) \,,
\nonumber \\
\phi_5^+ &=& - \frac{5}{6} \left(\lambda_1 (4 f_1^d - 1) + f_N( 3 +
4 V_1^d)\right)\,,
\nonumber \\
\psi_5^- &=& \frac{5}{3} \left(\lambda_1 (f_1^d - f_1^u) + f_N( 2 -
A_1^u - 3 V_1^d)\right)\,,
\nonumber \\
\psi_5^+ &=& -\frac{5}{6} \left(\lambda_1 (- 1 + 2 f_1^d +2 f_1^u) +
f_N( 5 + 2 A_1^u -2 V_1^d)\right)\,,
\nonumber \\
\xi_5^- &=& - \frac{5}{4} \lambda_2 f_2^d\,,
\nonumber \\
\xi_5^+ &=&  \frac{5}{36} \lambda_2 (2 - 9 f_2^d)\,,
\nonumber \\
\phi_6^- &=& \phantom{-}\frac{1}{2} \left(\lambda_1 (1- 4 f_1^d - 2
f_1^u) + f_N(1 +  4 A_1^u )\right) \,,
\nonumber \\
\phi_6^+ &=& - \frac{1}{2}\left(\lambda_1  (1 - 2 f_1^d) + f_N ( 4
V_1^d - 1)\right)\,. \nonumber
\end{eqnarray}

\end{document}